\RequirePackage{fix-cm}
\documentclass[smallextended]{svjour3}       
\smartqed  
\usepackage{booktabs} 
\usepackage{graphicx}
\usepackage{amsfonts}
\usepackage{dsfont}
\usepackage{graphicx, epsfig,amsmath,amssymb,latexsym,graphics,float}
\usepackage{setspace,subfig}
\usepackage{citesort}
\usepackage{float}
\usepackage{booktabs}
\usepackage{lipsum}
\usepackage{multicol}
\usepackage{url}
\usepackage{multirow}
\usepackage{algorithm}
\usepackage{algorithmic}
\begin{document}

\title{Robust Sequential Online Prediction with Dynamic Ensemble of Multiple Models: A Review
\thanks{This work was supported by Research Initiation Project (No.2021KB0PI01) and Exploratory Research Project (No.2022RC0AN02) of Zhejiang Lab.}
}

\author{Bin Liu
}

\institute{B. Liu \at
              Research Center for Applied Mathematics $\&$ Machine Intelligence, Zhejiang Lab, Hangzhou,
Zhejiang, 310021 China. \\
              \email{liubin@zhejianglab.com}
}

\date{Submitted to a Journal}

\maketitle

\begin{abstract}
The use of time series for sequential online prediction (SOP) has long been a research topic, but achieving robust and computationally efficient SOP with non-stationary time series remains a challenge. This paper reviews a framework, called Bayesian Dynamic Ensemble of Multiple Models (BDEMM), which addresses SOP in a theoretically elegant way, and have found widespread use in various fields. BDEMM utilizes a model pool of weighted candidate models, adapted online using Bayesian formalism to capture possible temporal evolutions of the data. This review comprehensively describes BDEMM from five perspectives: its theoretical foundations, algorithms, practical applications, connections to other research, and strengths, limitations, and potential future directions.
\end{abstract}
\section{Introduction}
\label{sec:introduction}
In one of his most influential papers \cite{breiman2001statistical}, Leo Breiman distinguishes between two cultures in statistical modeling: data modeling and algorithmic modeling. The former involves analyzing data based on a given stochastic model, such as linear or logistic regression, while the latter permits the data generative process to be unknown, and deploys models defined by algorithms. Modern discriminative deep neural networks belong to the algorithmic modeling culture and their architectures are often determined through the use of an architecture search algorithm \cite{elsken2019neural}. For more information on these modeling cultures, see \cite{breiman2001statistical}.

Leo Breiman's insights predate the rise of deep learning, yet he was a vocal proponent of the algorithmic modeling culture. He argued that overreliance on the data modeling culture could produce ``irrelevant theory, questionable conclusions" \cite{breiman2001statistical}.

We contend that the two cultures described above represent two extremes of a broader modeling culture, one that involves blending data with the modeler's prior knowledge. Under the data modeling culture, the model is predetermined by the modeler, based on their prior knowledge. In contrast, the algorithmic modeling culture yields a model by running an algorithm that the modeler has designed, incorporating their prior knowledge. Thus, the key discrepant factor between these two cultures pertains to how the prior knowledge of the modeler is deployed. For the data modeling culture, this knowledge is used directly in specifying the ``given stochastic data model", whereas for the algorithmic modeling culture, it is embedded within the algorithm employed to construct the model.

Apart from the two extreme cultures described above, there are other realizations of the broader modeling culture. The Bayesian Dynamic Ensemble of Multiple Models (BDEMM), which we discuss here, represents one such example. With BDEMM, a portion of the model structure is predetermined by the modeler, while the remaining part changes dynamically over time and is determined through an algorithm based on Bayesian inference.

BDEMM possesses three primary features, which render it appropriate for solving sequential online prediction (SOP) problems. These features are as follows: (1) BDEMM utilizes a model pool as opposed to a solitary model to account for the potential statistical patterns present in the data; (2) this model pool consists of several weighted candidate models, whose weights are adapted in real-time to reflect possible temporal fluctuations in the data; and (3) the method used to adjust the model weights is based on Bayesian formalism.

While time series have long been used for sequential online prediction (SOP), achieving robust and computationally efficient SOP with non-stationary time series remains a challenge \cite{liu2020sequential}. Traditional prediction methods train a model based on a static dataset and make predictions using the same model. This offline processing paradigm is unsuitable for SOP since it requires updating the model sequentially. In contrast, BDEMM addresses this issue by processing data sequentially. Upon the arrival of new data, it only updates the weights of its base models utilizing Bayesian formalism, while keeping the base models themselves unchanged. Unlike offline methods that require repeated access to data items, SOP with BDEMM necessitates only one pass through the data. Processed data items can be discarded immediately to free up storage space.

BDEMM has found widespread use in various fields, including dynamic system state tracking \cite{siebler2020joint,song2020particle,raftery2010online,liu2011instantaneous,dai2016robust,liu2017robust}, dynamic multi-modal data fusion \cite{liu2021robust}, Gaussian process-based online time-series prediction \cite{liu2020sequential}, Bayesian optimization \cite{liu2020harnessing}, and neural decoding for brain-computer interfaces \cite{qi2019dynamic}. In each of these contexts, BDEMM has provided a robust and state-of-the-art algorithmic solution. Despite its success in various applications, there has yet to be a comprehensive introduction to BDEMM. This serves as the motivation behind this introductory article.

This paper's contribution is that it provides, for the first time, a complete introduction to BDEMM, including its theoretical foundations, typical algorithms, practical applications, connections to other research, strengths, limitations, potential future directions, and additional available resources such as code and real-world data sets. These topics are discussed in Sections \ref{sec:theory} to \ref{sec:open}, respectively. The paper concludes in Section \ref{sec:conc}.
\section{Theories}\label{sec:theory}
This section introduces BDEMM from a theoretical perspective. We assume readers have prerequisite knowledge of Bayesian statistics, and some classic reference materials include Section IV of \cite{berger2013statistical}, Section II of \cite{barber2012bayesian}, and Section XI of \cite{wasserman2004all}. To begin, we provide a succinct overview of Bayesian model averaging (BMA), which serves as the theoretical foundation of BDEMM.
\subsection{Bayesian Model Averaging}
In traditional statistical modeling, a stochastic model $\mathcal{M}$ is initially specified to represent a hypothesis on how data are generated. With $\mathcal{M}$ possessing a fixed structure, the corresponding value of parameter $\theta$ is obtained through data fitting, which involves optimizing a criterion such as maximum likelihood (ML), maximum a posterior (MAP), or minimum mean square error (MMSE). One can also adhere to the Bayesian paradigm by assigning a prior density to $\theta$ and utilizing a likelihood function to characterize its probability given the data \cite{kruschke2018bayesian,kruschke2014doing,gelman1995bayesian,kruschke2010bayesian,berger2013statistical}. The ensuing task is to calculate the posterior of $\theta$, which is proportional to the product of the prior and the likelihood.

In many realistic scenarios, utilizing a model with an unchanging structure to capture the generative mechanism of the data is arbitrary. An alternate approach involves considering multiple candidate models, $\mathcal{M}_1, \mathcal{M}_2, \ldots, \mathcal{M}_K$, simultaneously to account for uncertainty at the model level \cite{peel2000finite,lesage1992mixture,wood2011bayesian}.

Given a set of independent identically distributed (i.i.d.) data points $\mathbf{y}=(y_1, \ldots, y_n)$, where $y_i\sim p(y_i|\theta_k)$, $i=1,\ldots,n$, where $\theta_k$ denotes the parameter of $\mathcal{M}_k$, the probability of $\mathbf{y}$ given $\mathcal{M}_k$ can be calculated as follows
\begin{equation}\label{eqn:marg_lik}
p_k(\mathbf{y})=\int_{\theta}p_k(\mathbf{y}|\theta)p_k(\theta)d\theta,
\end{equation}
where $p_k(\theta)$ and $p_k(\mathbf{y}|\theta)$ denote respectively the prior and the likelihood function defined in $\mathcal{M}_k$, $k=1,\ldots,K$. As is shown, $\theta$ is marginalized out (integrated out) in Eqn.(\ref{eqn:marg_lik}). Consequently, $p_k(\mathbf{y})$ is also referred to as the marginal likelihood or model evidence within the framework of Bayesian statistics.

Denoting the prior probability of $\mathcal{M}_k$ by $\mbox{Pr}(\mathcal{M}_k)$, we can obtain the posterior probability of $\mathcal{M}_k$ by applying Bayes formula
\begin{equation}\label{eqn:post_prob}
p(\mathcal{M}_k|\mathbf{y})=\frac{\mbox{Pr}(\mathcal{M}_k)p_k(\mathbf{y})}{\sum_{j=1}^{K}\mbox{Pr}(\mathcal{M}_j)p_j(\mathbf{y})}, k=1,\ldots, K.
\end{equation}

We denote the target parameter to be predicted as $\mathbf{x}$, with $\mathbf{x}\triangleq g(\theta)$ where $g$ is a function mapping $\theta$ to $\mathbf{x}$. The concept behind BMA is to take into account model uncertainty while predicting $\mathbf{x}$, as shown below \cite{fragoso2018bayesian,hoeting1999bayesian}
\begin{equation}\label{eqn:bma}
\hat{\mathbf{x}}=\sum_{k=1}^{K}\hat{\mathbf{x}}_kw_k,
\end{equation}
where $\hat{\mathbf{x}}_k$ denotes the estimated or predicted value of $x$ associated with $\mathcal{M}_k$ and $w_k\triangleq p(\mathcal{M}_k|\mathbf{y})$ the weight of $\mathcal{M}_k$ in making this estimation or prediction.
In the concept of BMA, selecting a model with an unchanging structure is abandoned by assigning probabilities to a set of candidate models: $\mbox{Pr}\{\mathcal{M}=\mathcal{M}_k\}=w_k, k=1,\ldots,K$.

For additional information on BMA, particularly its theoretical properties, we suggest referring to \cite{fragoso2018bayesian,hoeting1999bayesian}.
\subsection{Bayesian Dynamic Ensemble of Multiple Models}\label{sec:bdemm}
BMA does not involve a time variable and therefore cannot handle dynamic scenarios where data evolve over time. The BDEMM framework extends BMA to address dynamic systems in which the time variable plays a crucial role. In this setting, observations are denoted by $\mathbf{y}_t$, where $t\in\mathbb{R}$ with $t>0$ representing the discrete time index. The goal is to predict $x_t$ at every time step $t$ using data that have been collected up to time $t$, given as $\mathbf{y}_{1:t}\triangleq\{\mathbf{y}_{1},\ldots,\mathbf{y}_{t}\}$. Following Eqn.(\ref{eqn:bma}), the BMA equation for predicting $\mathbf{x}_t$ becomes:
\begin{equation}\label{eqn:bdemm}
\hat{\mathbf{x}}_t=\sum_{k=1}^{K}\hat{\mathbf{x}}_{k,t}w_{k,t},
\end{equation}
where $w_{k,t}\triangleq p(\mathcal{M}_k|\mathbf{y}_{1:t})$ and $\hat{\mathbf{x}}_{k,t}$ denotes the predicted value of $\mathbf{x}_t$ within $\mathcal{M}_k$.

BDEMM offers a general methodology for recursively computing Eqn.(\ref{eqn:bdemm}). In the BDEMM framework, each model component $\mathcal{M}_k$ comprises a state transition prior $p_k(\mathbf{x}_t|\mathbf{x}_{t-1})$ and a likelihood function $p_k(\mathbf{y}_t|\mathbf{x}_t)$. Given that $p_k(\mathbf{x}_{t-1}|\mathbf{y}_{1:t-1})$ is available, the predictive distribution of $\mathbf{x}_t$ associated with $\mathcal{M}_k$ is defined as
\begin{equation}\label{eqn:predictive}
p_k(\mathbf{x}_{t}|\mathbf{y}_{1:t-1})=\int_{\chi}p_k(\mathbf{x}_t|\mathbf{x}_{t-1})p_k(\mathbf{x}_{t-1}|\mathbf{y}_{1:t-1})d\mathbf{x}_{t-1},
\end{equation}
where $\chi$ denotes the value space of $\mathbf{x}$.
The posterior distribution of $\mathbf{x}_t$ linked to $\mathcal{M}_k$ is derived accordingly:
\begin{equation}\label{eqn:posterior}
p_k(\mathbf{x}_{t}|\mathbf{y}_{1:t})=\frac{p_k(\mathbf{y}_t|\mathbf{x}_t)p_k(\mathbf{x}_{t}|\mathbf{y}_{1:t-1})}{p(\mathcal{M}_k|\mathbf{y}_{1:t})},
\end{equation}
with the marginal likelihood (normalizing constant) of $\mathcal{M}_k$ being defined as:
\begin{equation}\label{eqn:Mk_evidence}
p(\mathcal{M}_k|\mathbf{y}_{1:t})=\int_{\chi}p_k(\mathbf{y}_t|\mathbf{x}_t)p_k(\mathbf{x}_{t}|\mathbf{y}_{1:t-1})d\mathbf{x}_{t}
\end{equation}
However, $p_k(\mathbf{x}_{t}|\mathbf{y}_{1:t})$ may not be analytically obtained. Monte Carlo methods like importance sampling \cite{bugallo2015adaptive,tokdar2010importance} or Markov Chain Monte Carlo (MCMC) \cite{brooks1998markov,gilks1995markov} can help draw random samples from it. The predicted value of $\mathbf{x}_{t}$ for $\mathcal{M}_k$, denoted by $\hat{\mathbf{x}}_{k,t}$, can be estimated via a criterion like ML, MMSE or MAP based on these samples.

In BDEMM, the recursive calculation of $w_{k,t}$ involved in Eqn.(\ref{eqn:bdemm}) begins by defining a Weight-Temporal-Transition (WTT) operator, which provides the predictive distribution of the correct model as follows:
\begin{equation}\label{eqn:predictive_model_weight}
w_{k,t|t-1}\triangleq \mbox{Pr}\{\mathcal{M}_{\mbox{correct},t}=\mathcal{M}_k|\mathbf{y}_{1:t-1}\}=f(w_{1:K,t-1}).
\end{equation}
Here $\mathcal{M}_{\mbox{correct},t}$ represents the correct model at time $t$, $w_{1:K,t-1}\triangleq\{w_{1,t-1},\ldots,w_{K,t-1}\}$. Potential options for the WTT operator are addressed in subsection \ref{sec:wtt}.

Once $w_{k,t|t-1}$ for $k=1,\ldots,K$ are specified, the Bayesian theorem can be applied to obtain $w_{k,t}$:
\begin{equation}\label{eqn:posterior_model_weight}
w_{k,t}=\frac{w_{k,t|t-1}p(\mathcal{M}_k|\mathbf{y}_{1:t})}{\sum_{j=1}^Kw_{j,t|t-1}p(\mathcal{M}_j|\mathbf{y}_{1:t})}, k=1,\ldots,K.
\end{equation}

In summary, starting with $p_k(x_{t-1}|\mathbf{y}_{1:t-1})$ and $w_{k,t-1}$ for $k=1,\ldots,K$, one can use Eqns.(\ref{eqn:predictive})-(\ref{eqn:posterior_model_weight}) to calculate $p_k(\mathbf{x}_{t}|\mathbf{y}_{1:t})$ as well as $w_{k,t}$ for $k=1,\ldots,K$. Usually, $p_k(\mathbf{x}_{t}|\mathbf{y}_{1:t})$ does not have an analytical form. Monte Carlo methods are often used to estimate $\hat{\mathbf{x}}_{k,t}$ based on samples drawn from $p_k(\mathbf{x}_{t}|\mathbf{y}_{1:t})$. Subsequently, after obtaining $\hat{\mathbf{x}}_{k,t}$ and $w_{k,t}$ for $k=1,\ldots,K$, $\hat{\mathbf{x}}_t$ can be derived using Eqn.(\ref{eqn:bdemm}).

\begin{remark}
BDEMM comprises a collection of BMA phases executed sequentially. A WTT operation is performed between each adjacent pair of BMA phases, generating the prior distribution of the correct model necessary for use in the subsequent BMA phase. If the WTT operation is appropriately chosen, the BDEMM framework inherits the desirable theoretical characteristics of BMA. With the help of the WTT operator, BDEMM allows BMA to be used in dynamic scenarios. In other words, BDEMM serves as a dynamic extension of the BMA theory.
\end{remark}
\subsection{WTT Operators}\label{sec:wtt}
As previously mentioned, BDEMM employs a WTT operator to provide the predictive distribution of the correct model at each time step. The following subsections review five practical options for the WTT operator.
\subsubsection{Operator I}\label{sec:wtt_I}
One simple choice for the WTT operator is to specify:
\begin{equation}\label{eqn:choice_I}
w_{k,t|t-1}=w_{k,t-1}, k=1,\ldots,K.
\end{equation}
This selection assumes that the model switching process is solely driven by the observed data, regardless of any prior knowledge of the modeler. When BDEMM is used to choose a single correct model from several candidate models to represent a stationary dataset whose data points come in a sequence, this operator is preferable over the others discussed below.
\subsubsection{Operator II}\label{sec:wtt_II}
Another straightforward WTT operator is to set:
\begin{equation}\label{eqn:choice_I}
w_{k,t|t-1}=C_k, k=1,\ldots,K,
\end{equation}
where $C_k, k=1,\ldots,K$ are constants specified by the modeler and satisfy $\sum_{k=1}^{K}C_k=1$. This option presumes that the prior knowledge represented by $C_k, k=1,\ldots,K$ thoroughly characterizes the model switching process. It is appropriate when the time-series observations provide no information for establishing the model switching law.
\subsubsection{Operator III}\label{sec:wtt_III}
The third WTT operator assumes that the transition of the correct model satisfies a Markov model defined by a $K$-by-$K$ mode transition matrix (MTM). An example of the MTM is shown below:
\begin{equation}\label{eqn:markov_matrix}
\mbox{T}=
\left(
  \begin{array}{cccc}
    0.9 & \frac{0.1}{K-1} & \dots & \frac{0.1}{K-1} \\
    \frac{0.1}{K-1} & 0.9 & \dots & \frac{0.1}{K-1} \\
    \vdots & \vdots & \vdots & \vdots \\
    \frac{0.1}{K-1} & \frac{0.1}{K-1} & \dots & 0.9 \\
  \end{array}
\right).
\end{equation}
Let the $\{i,j\}$th element of $\mbox{T}$ be denoted as $\mbox{T}_{ij}$. The value of $\mbox{T}_{ij}$ represents the probability that the correct model changes from $\mathcal{M}_i$ to $\mathcal{M}_j$, $i=1,\ldots,K, j=1,\ldots,K$, in a single time step. The definition of $\mbox{T}$ presented in Eqn.(\ref{eqn:markov_matrix}) implies an assumption that the correct model transitions infrequently (with a likelihood of 10\% per time instance). With $\mbox{T}$ fixed, the WTT operation defined by Eqn.(\ref{eqn:predictive_model_weight}) can be expressed as:
\begin{equation}\label{eqn:predictive_model_weight2}
w_{k,t|t-1}=\sum_{i=1}^Kw_{i,t-1}\mbox{T}_{ik}, k=1,\ldots,K.
\end{equation}
In contrast to the data-independent operator II, this operator is data-dependent. The prior model probability at time $t$ is related to its posterior probability at time $t-1$, where the latter is influenced by the data item collected at time $t-1$.
\subsubsection{Operator IV}\label{sec:wtt_IV}
As shown above, setting up a matrix such as Eqn.(\ref{eqn:markov_matrix}) necessitates assigning values for $K^2$ elements. Typically, there is insufficient information to accomplish this task. It is preferable to specify the MTM using a forgetting process, which only requires identifying one value for a hyperparameter called the forgetting factor. The concept of forgetting dates back to works by \cite{fagin1964recursive,jazwinski2007stochastic}.

Let the forgetting factor be denoted as $\alpha, 0<\alpha<1$. With this value, the WTT operation can be defined as:
\begin{equation}\label{eqn:predictive_model_weight4}
w_{k,t|t-1}=\frac{w_{k,t-1}^{\alpha}}{\sum_{i=1}^{K}w_{i,t-1}^{\alpha}}, k=1,\ldots,K.
\end{equation}
This forgetting-based approach is often utilized in the BDEMM framework. An empirical rule for selecting the value of $\alpha$ is to choose a value slightly below 1, indicating the assumption that the model switching process is smooth. This operator is more practical than operator III when one knows that the model transition law corresponds to a Markov chain but lacks adequate prior information to define an appropriate MTM.
\subsubsection{Operator V}\label{sec:wtt_V}
The final WTT operator presumes that the model-switching process can be described or approximated by a P$\acute{o}$lya urn process, as proposed in \cite{el2021particle}. It sets:
\begin{equation}\label{eqn:predictive_model_weight5}
w_{k,t|t-1}=\frac{\beta_k+\sum_{\tau=1}^{t-1}w_{k,\tau}}{\sum_{j=1}^{K}\left(\beta_j+\sum_{\tau=1}^{t-1}w_{j,\tau}\right)}, k=1,\ldots,K,
\end{equation}
where $\beta_k\in\mathbb{N}$ is a positive integer predetermined by the modeler for $k=1,\ldots,K$. As per Eqn.(\ref{eqn:predictive_model_weight5}), the probability of transitioning to a particular model at a given time step depends on all prior weights received by that model. It has been demonstrated that breaking the Markovian switching assumption and exploiting long-term memory for the correct model can improve performance for some applications \cite{el2021particle}. This operator outperforms the other methods when the model-switching process lacks Markov property but possesses a long-term memory structure approximated by a P$\acute{o}$lya urn process.
\subsection{Importance Sampling for Marginal Likelihood Estimation}
To implement the BDEMM framework in practice, a crucial computation challenge involves evaluating the marginal likelihood, i.e., Eqn.(\ref{eqn:Mk_evidence}). The stochastic integral in Eqn.(\ref{eqn:Mk_evidence}) may be amenable to analytical solution for some exceptional scenarios where the prior and likelihood functions are conjugate. For most practical cases, however, no such exact analytical solutions exist. Accordingly, BDEMM leverages importance sampling to generate an approximation of the marginal likelihood.

In the following, we denote the product of the prior and likelihood by $\pi(\mathbf{x})$, which is proportional to the posterior associated with our target model $\mathcal{M}$. The importance sampling method for estimating the marginal likelihood starts by specifying a proposal distribution $q(\mathbf{x})$ that is absolutely continuous with respect to $\pi(\mathbf{x})$. With $q(\mathbf{x})$ defined, the marginal likelihood of $\mathcal{M}$ can be expressed as \cite{geweke1989bayesian,oh1992adaptive}:
\begin{equation}\label{eqn:is_for_si}
l(\mathcal{M})=\mathbb{E}_q\left[\frac{\pi(\mathbf{x})}{q(\mathbf{x})}\right],
\end{equation}
where $\mathbb{E}_q$ denotes the expectation operation with respect to distribution $q$. Drawing a set of i.i.d. random samples $x^1,\ldots,x^N$ drawn from $q$, an unbiased and consistent Monte Carlo estimate of the marginal likelihood can be formulated as:
\begin{equation}\label{eqn:is_ml}
\hat{l}(\mathcal{M})=\frac{1}{N}\sum_{i=1}^{N}\frac{\pi(x^i)}{q(x^i)},
\end{equation}
where $\frac{\pi(x^i)}{q(x^i)}$ is referred to as the importance weight of the $i$th sample.

The efficiency of the estimate given in Eqn.(\ref{eqn:is_ml}) hinges on the choice of the proposal distribution $q$ \cite{kong1994sequential}. A simple and direct option is to set $q$ equal to the prior; then, the estimator becomes the average of the likelihoods of the samples as the prior terms cancel out in the numerator and denominator. An empirical rule of thumb for selecting $q$ is to design it to approximate the posterior's shape as closely as possible \cite{oh1992adaptive}. To this end, adaptive importance sampling (AIS) techniques have been developed that iteratively fine-tune $q$ in a data-driven fashion \cite{oh1992adaptive,cappe2008adaptive,bugallo2017adaptive}. In situations where the posterior is highly multimodal, the annealing strategy can be incorporated into AIS, resulting in the adaptive annealed importance sampling (AAIS) algorithm \cite{liu2014adaptive}.
\section{Algorithms}\label{sec:alg}
We will now present the three major algorithms developed in the BDEMM framework. The algorithm's form relies on the model structures, such as the state transition prior and likelihood function. If they are both Gaussian, then Kalman filtering (KF) is the most suitable choice to compute Eqns.(\ref{eqn:predictive})-(\ref{eqn:Mk_evidence}) \cite{welch1995introduction,maybeck1990kalman}. For cases involving nonlinear and/or non-Gaussian systems, the Sequential Monte Carlo (SMC) methodology is often utilized to provide an approximate estimate of the true answer \cite{liu1998sequential,doucet2001sequential}. In addition to the KF- and SMC-based BDEMM approaches, we also describe an algorithm known as INstant TEmporal structure Learning (INTEL) \cite{liu2020sequential}. The INTEL approach can be viewed as a Gaussian process time series (GPTS) model-based implementation of the BDEMM theory. In each algorithm, one WTT operator from those described in subsection \ref{sec:wtt} can be selected for use.
\subsection{KF-based BDEMM}\label{sec:kf_bdemm}
Let us assume that the state transition prior and the likelihood function for time instance $t$ under $\mathcal{M}_k$ are denoted by $\mathcal{N}(\mathbf{A}_k\mathbf{x}_{t-1},\mathbf{Q}_k)$ and $\mathcal{N}(\mathbf{B}_k\mathbf{x}_{t},\mathbf{R}_k)$, respectively. Here $\mathcal{N}(\mu,\Sigma)$ denotes a Gaussian distribution with mean $\mu$ and covariance $\Sigma$.

Given $p_k(\mathbf{x}_{t-1}|\mathbf{y}_{1:t-1})=\mathcal{N}(\hat{\mathbf{x}}_{k,t-1},\Sigma_{k,t-1}), k=1,\ldots, K$, we can obtain the predictive distribution of the state using Eqn.(\ref{eqn:predictive}), as per the BDEMM theory presented in Subsection \ref{sec:bdemm}. For the state transition prior and likelihood function specified above, Eqn.(\ref{eqn:predictive}) simplifies to \cite{welch1995introduction,maybeck1990kalman}
\begin{equation}\label{eqn:KF_predict}
p_k(\mathbf{x}_{t}|\mathbf{y}_{1:t-1})=\mathcal{N}(\hat{\mathbf{x}}_{k,t|t-1},\mathbf{P}_{k,t|t-1}),
\end{equation}
where $\hat{\mathbf{x}}_{k,t|t-1}=\mathbf{A}_k\hat{\mathbf{x}}_{k,t-1}$, $\mathbf{P}_{k,t|t-1}=\mathbf{A}_k\Sigma_{k,t-1}\mathbf{A}_k^T+\mathbf{Q}_k$, and $\mathbf{A}^T$ represents the transpose of $\mathbf{A}$.

According to the BDEMM theory presented in Subsection \ref{sec:bdemm}, we can obtain the posterior distribution of $x_t$ associated with $\mathcal{M}_k$ by computing Eqn.(\ref{eqn:posterior}), which translates to \cite{welch1995introduction,maybeck1990kalman}
\begin{equation}\label{eqn:kf_update}
p_k(\mathbf{x}_{t}|\mathbf{y}_{1:t})=\mathcal{N}(\hat{\mathbf{x}}_{k,t|t-1}+\mathbf{G}_{k,t}\tilde{\mathbf{Z}}_t,\mathbf{P}_{k,t|t}),
\end{equation}
where $\mathbf{G}_{k,t}=\mathbf{P}_{k,t|t-1}\mathbf{B}_k^T\mathbf{S}_{k,t}^{-1}$, $\tilde{\mathbf{Z}}_t=\mathbf{y}_t-\mathbf{B}_k\hat{\mathbf{x}}_{k,t|t-1}$, $\mathbf{A}^{-1}$ denotes the inverse of $\mathbf{A}$, $\mathbf{S}_{k,t}=\mathbf{B}_k\mathbf{P}_{k,t|t-1}\mathbf{B}_k^{T}+\mathbf{R}_k$, and $\mathbf{P}_{k,t|t}=\mathbf{P}_{k,t|t-1}-\mathbf{G}_{k,t}\mathbf{B}_k\mathbf{P}_{k,t|t-1}$.

In the above scenario, $p_k(\mathbf{x}_{t}|\mathbf{y}_{1:t-1})$ and the likelihood function are both Gaussian. Thus, the marginal likelihood specified by Eqn.(\ref{eqn:Mk_evidence}) can be computed as an integral of the product of two Gaussians, which can be solved numerically.

Given a WTT operator as described in Subsection \ref{sec:wtt}, one can compute $w_{k,t|t-1}$ using Eqn.(\ref{eqn:predictive_model_weight}) and then obtain $w_{k,t}$ using Eqn.(\ref{eqn:posterior_model_weight}). We can then use the set of $w_{k,t}, k=1,\ldots,K$ to determine an average mean and covariance for use in the next time step. This method approximates the Gaussian mixture $\sum_{k=1}^Kw_{k,t}p_k(\mathbf{x}_{t}|\mathbf{y}_{1:t})$ with a single Gaussian, which we employ to initialize the $K$ Gaussian components for the next time step. It is important to note that if most of the Gaussian components in the mixture distribution possess negligibly small weights, then the mixture distribution can be accurately represented by a single Gaussian. However, if the majority of the Gaussian components have significant weights, approximation errors may occur. The main concern when approximating the Gaussian mixture with a single Gaussian is computational efficiency. Employing this technique enables us to limit the number of Gaussian components in the posterior from rapidly growing over time.
\subsection{SMC-based BDEMM}\label{sec:smc_bdemm}
The SMC method employs a set of weighted samples to approximate a sequence of target distributions. Unlike KF, SMC does not restrict the model to be linear Gaussian. Assume that the posterior distribution of $\mathbf{x}_{t-1}$ is approximated by a set of weighted samples $\{x_{t-1}^i,u_{t-1}^i\}_{i=1}^N$, where $N$ denotes the sample size. This can be written as
\begin{equation}\label{eqn:sample_approx_1}
p_k(\mathbf{x}_{t-1}|\mathbf{y}_{1:t-1})\simeq\sum_{i=1}^{N}u_{t-1}^i\delta(\mathbf{x}_{t-1}-x_{t-1}^i),
\end{equation}
where $\delta(x)$ takes a value of 1 if $x=0$ and 0 otherwise.
Next, we describe how to obtain $w_{k,t}$ and the sample set $\{x_{t}^i,u_{t}^i\}_{i=1}^N$ for approximating $p(\mathbf{x}_{t}|\mathbf{y}_{1:t})$, based on $\{x_{t-1}^i,u_{t-1}^i\}_{i=1}^N$ and $w_{k,t-1}$, $k=1,\ldots,K$, following theories presented in Section \ref{sec:theory}.

Given $\{x_{t-1}^i,u_{t-1}^i\}_{i=1}^N$ and $p_{k}\left(\mathbf{x}_{t} \mid \mathbf{x}_{t-1}\right)$, one can generate a new sample set $\{x_{k,t}^i,u_{t-1}^i\}_{i=1}^N$, in which $x_{k,t}^i$ is drawn from $p_{k}\left(\mathbf{x}_{t} \mid x_{t-1}^i\right)$. Then the predictive distribution $p_k(\mathbf{x}_{t}|\mathbf{y}_{1:t-1})$ given by Eqn.(\ref{eqn:predictive}) in BDEMM theory can be approximated as follows:
\begin{equation}\label{eqn:sample_approx_2}
p_k(\mathbf{x}_{t}|\mathbf{y}_{1:t-1})\simeq\sum_{i=1}^{N}u_{t-1}^i\delta(\mathbf{x}_{t}-x_{k,t}^i).
\end{equation}

According to the BDEMM theory presented in Subsection \ref{sec:bdemm}, the next step is to compute the posterior distribution of $x_t$ associated with $\mathcal{M}_k$ using Eqn.(\ref{eqn:posterior}). Due to the nonlinearity and non-Gaussianity of the models, this posterior distribution does not have an analytical solution. Therefore, we use the importance sampling technique to approximate it. Specifically, we update the sample weight as follows:
\begin{eqnarray}\label{eqn:weight_update}
\tilde{u}_{k,t}^i&=&u_{t-1}^ip_{k}\left(\mathbf{y}_{t} \mid x_{k,t}^i\right), i=1,\ldots,N, \\
u_{k,t}^i &=& \frac{\tilde{u}_{k,t}^i}{\sum_{j=1}^{N}\tilde{u}_{k,t}^j}, i=1,\ldots,N,
\end{eqnarray}
Then, in accordance with the importance sampling theory \cite{tokdar2010importance}, the updated sample set $\{x_{k,t}^i,u_{k,t}^i\}_{i=1}^N$ constitutes a Monte Carlo approximation to $p_k(\mathbf{x}_{t}|\mathbf{y}_{1:t})$, which can be expressed as follows:
\begin{equation}\label{eqn:sample_approx_3}
p_k(\mathbf{x}_{t}|\mathbf{y}_{1:t})\simeq\sum_{i=1}^{N}u_{k,t}^i\delta(\mathbf{x}_{t}-x_{k,t}^i).
\end{equation}
Finally, the estimation of $\hat{\mathbf{x}}_{k,t}$ can be obtained from the sample set $\{x_{k,t}^i,u_{k,t}^i\}_{i=1}^N$ based on a specified criterion, such as ML, MAP, or MMSE.

Given $w_{k,t-1}$, one can choose a WTT operator (see Subsection \ref{sec:wtt}) to determine $w_{k,t|t-1}$, $k=1,\ldots,K$. By substituting Eqn.(\ref{eqn:sample_approx_2}) into Eqn.(\ref{eqn:Mk_evidence}), we obtain a Monte Carlo estimate of the marginal likelihood as follows:
\begin{equation}\label{eqn:sample_approx_4}
p\left(\mathcal{M}_{k} \mid \mathbf{y}_{1: t}\right)\simeq\sum_{i=1}^Nu_{t-1}^ip_{k}\left(\mathbf{y}_{t} \mid x_{k,t}^i\right), k=1,\ldots,K.
\end{equation}
Next, we substitute $w_{k,t|t-1}$ and Eqn.(\ref{eqn:sample_approx_4}) into Eqn.(\ref{eqn:posterior_model_weight}), resulting in the computation of $w_{k,t}$ for $k=1,\ldots,K$. Finally, the estimation of $\mathbf{x}_t$ can be obtained using Eqn.(\ref{eqn:bdemm}).

In the context of BDEMM, the posterior has a mixture form given by:
\begin{equation}\label{eqn:mixture_form}
p(\mathbf{x}_{t}|\mathbf{y}_{1:t})=\sum_{k=1}^{K}w_{k,t}p_k(\mathbf{x}_{t}|\mathbf{y}_{1:t}).
\end{equation}
By substituting Eqn.(\ref{eqn:sample_approx_3}) into the above equation, we obtain a Monte Carlo approximation to the posterior:
\begin{equation}\label{eqn:sample_approx_5}
p(\mathbf{x}_{t}|\mathbf{y}_{1:t})\simeq\sum_{k=1}^{K}\sum_{i=1}^{N}w_{k,t}u_{k,t}^i\delta(\mathbf{x}_{t}-x_{k,t}^i).
\end{equation}

To overcome particle degeneracy \cite{arulampalam2002tutorial}, a resampling procedure is used to draw a set of equally weighted samples from the augmented sample set as follows:
\begin{equation*}
\mathbf{D}\triangleq\{\{x_{1,t}^i, w_{1,t}u_{1,t}^i\}_{i=1}^N,\ldots,\{x_{K,t}^i, w_{K,t}u_{K,t}^i\}_{i=1}^N\}.
\end{equation*}
This set is then used to approximate $p(\mathbf{x}_{t}|\mathbf{y}_{1:t})$ in Eqn.(\ref{eqn:sample_approx_5}). Let us denote $\mathbf{D}$ as $\mathbf{D}\triangleq\{\bar{x}_t^i, \bar{u}^i\}_{i=1}^{NK}$ for simplicity in notation.
The corresponding resampling process involves the following steps, which eventually generate an equally weighted sample set $\{x_t^n, 1/N\}_{n=1}^{N}$.
\begin{itemize}
\item For $\forall n\in\{1,\ldots,N\}$:
\begin{enumerate}
\item Draw a random sample $v$ from the uniform distribution $\mathcal{U}(0,1)$.
\item Identify an index $j$ such that $\sum_{i=1}^j\bar{u}^i\leq
v\leq\sum_{i=1}^{j+1}\bar{u}^i$.
\item Set $x_t^n=\bar{x}_t^j$.
\end{enumerate}
\end{itemize}
Readers can refer to \cite{Hol2006on} for more details on resampling procedures that are commonly employed in SMC.

The updated sample set $\{x_t^i, 1/N\}_{i=1}^{N}$ along with the model weights $w_{k,t}, k=1,\ldots,K$ will be utilized in the subsequent time instance $t+1$ of the algorithm iteration.

It is important to note that the aforementioned procedure presents a general version of the implementation of SMC-based BDEMM. In certain instances, the implementation can be simplified. For instance, when all models share the same state transition prior, as observed in \cite{liu2021robust,liu2017robust}, we only need to generate a single updated sample set $\{x_t^i\}_{i=1}^N$ to approximate the posterior and compute the marginal likelihood for every model. This approach reduces the memory and computational requirements significantly.
\subsection{GPTS-based BDEMM}\label{sec:intel}
The GPTS-based BDEMM method, also referred to as the INstant TEmporal structure Learning (INTEL) algorithm in \cite{liu2020sequential}, integrates the concepts of GPTS modeling and BDEMM. It utilizes multiple weighted GPTS models simultaneously. The model weights are updated in the same manner as described in Eqns.(\ref{eqn:predictive_model_weight})-(\ref{eqn:posterior_model_weight}). Nonetheless, instead of point predictions, the model averaging is performed based on predictive distributions. Specifically, each GPTS model generates a predictive Gaussian distribution of $\mathbf{x}_t$. A weighted generalization of the product of experts (POE) approach is then utilized to combine these Gaussian distributions and produce the final prediction.
In the context of INTEL, the objective to be predicted at time step $t$, i.e. $\mathbf{x}_t$, corresponds to the upcoming data point for the subsequent time step, which is denoted by $\mathbf{y}_{t+1}$.

Due to the elegant theoretical properties of Gaussian processes (GPs) (see \cite{williams2006gaussian} for details), for each GPTS model, denoted as $\mathcal{M}_k$, one can directly and analytically compute its marginal likelihood $p(\mathcal{M}_k|\mathbf{y}_{1:t})$ without resorting to the definition of state transition prior and likelihood function. Specifically, the GPTS model posits that the observation $\mathbf{y}_t$ follows a Gaussian distribution, which is given by \cite{roberts2013gaussian}:
\begin{equation}
\mathbf{y}_{t}=f(t)+\eta_{t},
\end{equation}
where $f \sim \mathcal{GP}\left(\mu, k_{\theta}\right), \eta_{t} \sim \mathcal{N}\left(0, \sigma_n^{2}\right)$, $\mathcal{GP}\left(\mu, k_{\theta}\right)$ represents a GP with a mean function $\mu(\cdot)$ and a covariance kernel function $k_{\theta}(\cdot,\cdot)$ parameterized with $\theta$.
Then, based on the historical dataset $\{\mathbf{t},\mathbf{y}\}$ where $\mathbf{t}=\{t-\tau+1,\ldots,t\}$ and $\mathbf{y}=\{\mathbf{y}_{t-\tau+1},\ldots,\mathbf{y}_t\}$, we can derive that the predictive distribution of $y_{t+1}$ is Gaussian with an analytically solvable mean and variance. Here $\tau$ denotes the length of the considered time window.

The fundamental concept underlying the INTEL algorithm is to utilize multiple GPTS models together, each capturing one type of temporal structure of the data by using a set of hyper-parameter values. Let $p_k(\mathbf{y}_{t+1})$ be the predictive distribution of $\mathbf{y}_{t+1}$ generated by $\mathcal{M}_k$. The final predictive distribution provided by the INTEL algorithm is represented as:
\begin{equation}\label{eqn:generalized_poe}
p(\mathbf{y}_{t+1})\propto \Pi_{k=1}^K \left(p_k(\mathbf{y}_{t+1})\right)^{\omega_{k,t+1|t}},
\end{equation}
where $\omega_{k,t+1|t}$ is given by the WTT operation (refer to subsection \ref{sec:wtt} for details). It is worth noting that $p(\mathbf{y}_{t+1})$ is also Gaussian with an analytically solvable mean and variance. For more information on the INTEL algorithm, readers can refer to \cite{liu2020sequential}.
\section{Applications}\label{sec:app}
This section outlines various classical applications of BDEMM, which demonstrate its versatility.
\subsection{Robust Tracking of Dynamic System States}
\subsubsection{Robust SMC methods}\label{sec:rpf}
SMC methods, also known as particle filters (PFs), have been widely used in nonlinear non-Gaussian state filtering. However, their performance may deteriorate rapidly when outliers appear intermittently in the measurements due to sudden changes in the environment such as sensor faults \cite{liu2015toward,liu2017state,wang2017online}. To handle these outliers effectively and avoid divergence of PFs, a robust PF (RPF) algorithm has been proposed in \cite{liu2017robust}, which is a type of SMC-based BDEMM method. The RPF algorithm incorporates a nominal Gaussian noise model alongside two heavy-tailed Student's t noise models into the model pool to account for both regular observations and outlier-contaminated ones. Using the dynamic model re-weighting mechanism provided by BDEMM, the nominal model dominates when regular observations arrive, while the one heavy-tailed Student's t model takes over and dominates the filtering process when an outlier arrives. Therefore, the influence of the outlier on the filtering performance is mitigated automatically.

The BDEMM-type SMC methods have found significant applications across various domains including \cite{siebler2020joint,de2016automatic,song2020particle,liu2011instantaneous,dai2016robust}. For instance, a research team from German Aerospace Center and Karlsruhe Institute of Technology adapted the RPF method to track a high-speed moving train utilizing earth magnetic field distortion data \cite{siebler2020joint}. A positive experimental outcome was reported, indicating that the RPF algorithm is robust against periodic noise caused by currents in the overhead line, distortions introduced by passing trains or outdated values in the map of the magnetic field along the railway tracks. Using RPF, they achieved an overall performance equivalent to a root-mean-square-error (RMSE) below five meters with only the earth magnetic field distortion data. This performance is comparable to those obtained with the global navigation satellite system (GNSS).

In \cite{liu2011instantaneous}, a BDEMM-type SMC method was employed for tracking the instantaneous frequency of a non-stationary signal. The challenge in this task lies in estimating the instantaneous frequency accurately online, while the instantaneous frequency may vary over time irregularly. In the proposed BDEMM-type SMC algorithm, six plausible instantaneous frequency evolution models are considered together, and the BDEMM theory is adopted to address the uncertainty on which model to use at each time instance. These candidate models have different state-transition priors but share the same nonlinear likelihood function. The proposed algorithm employs a forgetting-based WTT operator and achieved remarkable performance consistently over various cases.

In \cite{dai2016robust}, an SMC-based BDEMM type method was introduced for tracking a moving object in a low-resolution video stream. Two candidate models were used, corresponding to the color feature and the texture feature of the object to be tracked. They shared the same state-transition prior that captures the temporal correlation in the sequence of frame images. The difference between the candidate models lies in the likelihood function as they utilize different features. Using the BDEMM theory, one feature takes the larger effect when the other feature changes abruptly due to occlusion or the appearance of confusing colors. The proposed algorithm demonstrated outstanding performance in terms of robustness, expressivity, and comprehensibility.
\subsubsection{Robust KF}
In \cite{raftery2010online}, the authors address the task of accurately predicting online output strip thickness for a cold rolling mill. Several physically motivated models are available, but it is uncertain which one should be used. A Markov chain model, given in terms of forgetting, is utilized to capture the temporal transition of the correct model. The resulting method, called dynamic model averaging (DMA), is KF-based BDEMM in spirit. The authors demonstrate that DMA outperforms the best physical model and quickly converges to it when it is included in the model pool.
\subsection{Dynamic Multi-Modal Data Fusion}\label{sec:mmdf}
From the viewpoint of model uncertainty, a type of SMC-based BDEMM algorithm is introduced for robust multi-modal data fusion (RMMDF) in dynamic systems \cite{liu2021robust}. The RMMDF approach utilizes a set of candidate models, each representing a hypothesis on modality failure (or failures). When a modality failure arises, the RMMDF approach quickly down-weights the candidate models connected with the failing modality (or modalities) and simultaneously enhances the weights of other candidate models. This behavior is due to the Bayesian formulation of BDEMM.
\subsection{Robust GPTS based Online Prediction}
In time-series online prediction, the presence of outliers and/or change points is a significant obstacle to overcome, particularly when the temporal structure of the data evolves over time. To address this challenge, a robust online prediction algorithm has been proposed in \cite{liu2020sequential}. This algorithm utilizes the BDEMM framework and embeds multiple GPTS models, each representing a different type of temporal structure. The main part of the resulting algorithm, called INTEL, is briefly introduced in subsection \ref{sec:intel}. The robustness of the algorithm is demonstrated by extensive experimental results utilizing real data, as detailed in \cite{liu2020sequential}.
\subsection{Bayesian Optimization}
To accelerate the search for the global optimum of a black-box objective function that is expensive to evaluate using low-fidelity (LF) data, an INTEL-type BDEMM algorithm called accelerated Bayesian Optimization (ABO) has been proposed in \cite{liu2020harnessing}. ABO dynamically weights two candidate models: a LF GP model and a high-fidelity (HF) GP model, similar to INTEL. During the initial searching phase when less HF data are available, the weight of the LF GP model takes on a larger value. As more HF data become available and are used to update the HF GP model, the LF GP model is gradually down-weighted following Bayesian formalism. Consequently, ABO offers an elegant solution to leveraging LF data to accelerate BO searching processes without compromising search quality.
\subsection{Neural Decoding for Brain-Computer Interfaces}
For intracortical brain-computer interfaces, a non-stationary neural decoding approach called dynamic ensemble modeling (DyEnsemble) has been proposed in \cite{qi2019dynamic}. Unlike other prevalent neural decoders, which assume a static mapping relationship from neural signals to motor intention, DyEnsemble permits this mapping relationship to vary over time by utilizing diverse measurement models that are dynamically weighted. The authors demonstrate that DyEnsemble is remarkably effective at mitigating the degradation of decoding performance caused by noise, missing neural data, or changes in brain activities due to neuroplasticity. The decoding process of DyEnsemble is similar to an SMC-based BDEMM procedure. Compared to other BDEMM methods, DyEnsemble stands out for its candidate model generation strategy, which was inspired by the Dropout operation originally developed in the context of deep artificial neural networks \cite{srivastava2014dropout}.
\subsection{A Toy Experiment}\label{sec:toy_exp}
Consider the time-series experiment presented in \cite{van2000the}. The hidden state $\mathbf{x}$ underlying the time-series observations $\mathbf{y}$ evolves over time, given by
\begin{equation}
\mathbf{x}_{t+1}=1+\sin(0.04\pi\times(t+1))+0.5\mathbf{x}_t+\mathbf{u}_t,
\end{equation}
where $\mathbf{u}_t$ is a Gamma(3,2) distributed random noise item. The observation at time $t$ is related to $\mathbf{x}_{t}$ as follows
\begin{equation}
\mathbf{y}_t=\left\{\begin{array}{ll}
0.2\mathbf{x}_t^2+\mathbf{n}_t,\quad\quad\quad t\leq30 \\
0.2\mathbf{x}_t-2+\mathbf{n}_t,\quad\, t>30 \end{array} \right.
\end{equation}
Here $\mathbf{n}_t$ represents the observation noise. The objective is to track the hidden state $\mathbf{x}_t$ in real-time based on noisy observations $\mathbf{y}_{1:t}$, $t=1,\ldots,60$. We showcase the basic features and advantages of BDEMM through this toy experiment, and the code is available at \url{https://github.com/robinlau1981/BDEMM}.

In the original experiment presented in \cite{van2000the}, $\mathbf{n}_t$ is drawn from \emph{a priori} known zero-mean Gaussian distribution.
In this more challenging case, the distribution of $\mathbf{n}_t$ is time-varying and drawn from a uniform distribution between 40 and 50 at specific ``abnormal observation times" or a zero-mean Gaussian distribution otherwise. We adapt the SMC-based BDEMM algorithm presented in subsection \ref{sec:smc_bdemm} to deal with this task. Two candidate models are employed: one assumes $\mathbf{n}_t$ is zero-mean Gaussian distributed, while the other assumes it is uniformly distributed between -50 and 50. We compare our algorithm with two single-model-based SMC methods, SMC-I and SMC-II, which assume $\mathbf{n}_t$ to be zero-mean Gaussian distributed and uniformly distributed between -50 and 50, respectively.

SMC-I and SMC-II's performance is expected to degrade when their model assumptions do not match the data generation process. However, the SMC-based BDEMM provides an automatic mechanism to switch between its two candidate models when the data regime changes. Specifically, when an ``abnormal observation time" occurs, the likelihood values of the state samples corresponding to the 2nd candidate model shall become significantly larger than those to the 1st one. Then, the marginal likelihood of the 2nd model shall become much larger, leading to an increase in the weight of the 2nd candidate model.

\begin{figure}[h]
\centering
\includegraphics[width=3.5in,height=2.1in]{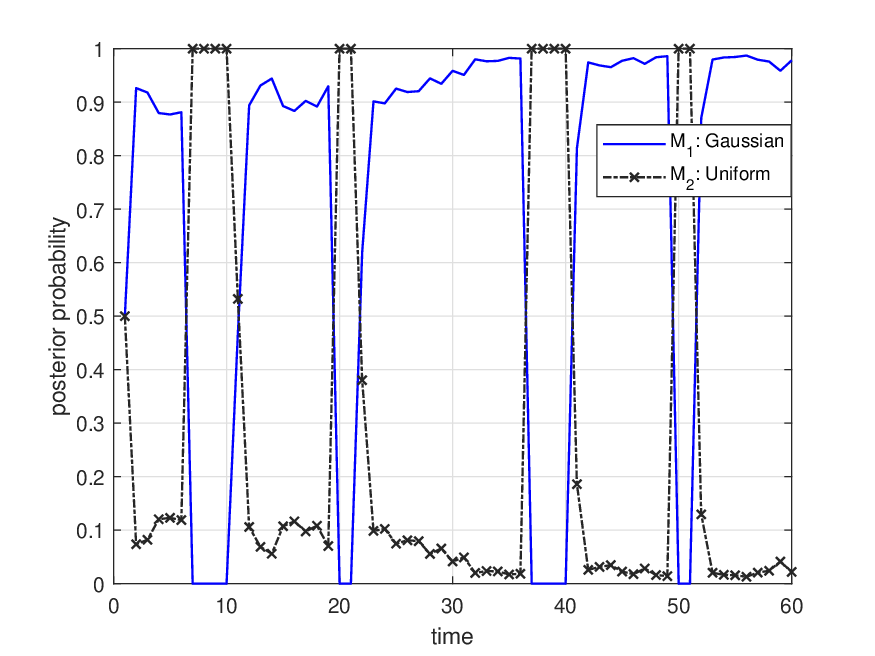}
\caption{Averaged posterior probabilities of the candidate models outputted by the SMC-based BDEMM method over 30 independent runs}\label{fig:model_prob}
\end{figure}

The weights or posterior probabilities of the member models averaged over 30 independent runs of our algorithm are plotted in Fig. \ref{fig:model_prob}. The weight of the 2nd model quickly increases when an ``abnormal observation time" arises, indicating that the algorithm detects the abnormality in the observations and selects the right model accordingly. Table 1 presents their performances in terms of the mean square error (MSE). Our BDEMM outperforms the others significantly.
\begin{table}[h]\centering\small
\caption{Mean and variance of the MSE calculated over 30 independent runs for each algorithm}
\begin{tabular}{c||c|c}
\hline %
 & \multicolumn{2}{c}{MSE}\\
&mean&var\\\hline
BDEMM &0.58675&0.016263\\\hline
SMC I&0.61869&0.025482\\\hline
SMC II&0.65352&0.032834\\\hline
\end{tabular}
\label{Table:performance}
\end{table}
\section{Connections to Related Research}\label{sec:connections}
\begin{figure}[h]
\centering
\includegraphics[width=3.6in,height=1.5in]{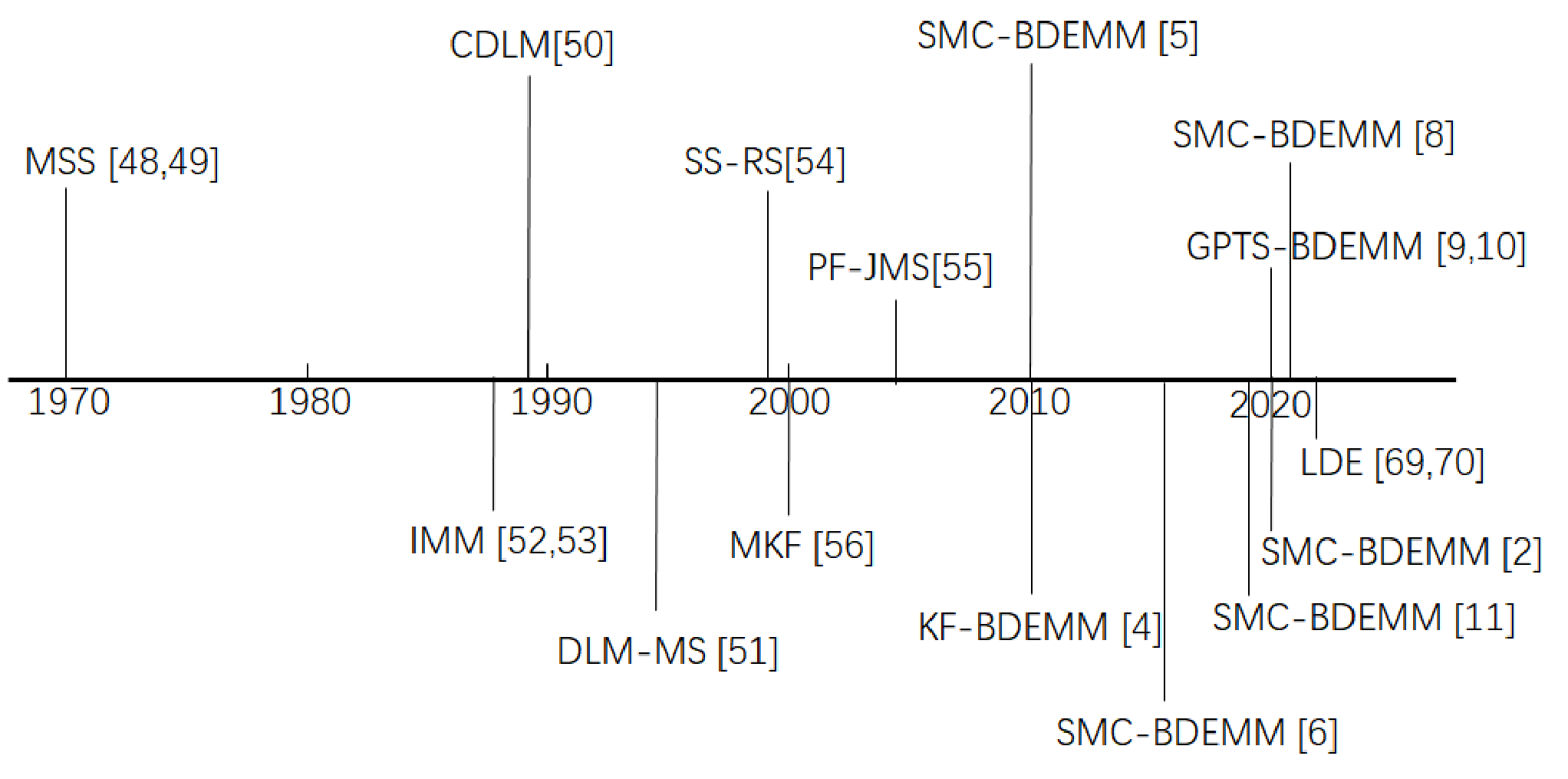}
\caption{A timeline of significant works related to BDEMM. The abbreviations used in this figure are defined as follows. MSS refers to Markov switching systems, which are models that allow the probability distribution of an observation to vary over time due to unobserved latent states. CDLM denotes conditional dynamic linear models, which employ models with changing parameters to deal with non-stationarity. IMM stands for interacting multiple models, which is another technique to handle non-stationarity through a mixture of models where the weights change depending on time or other factors. DLM-MS represents dynamic linear models with Markov-switching, a type of state space model that can represent non-stationary time series. SS-RS refers to state-space models with regime switching, a class of models that combine state-space modeling and Markov-switching ideas to deal with non-stationarity. MKF stands for mixture Kalman filters, which utilize the Kalman filter framework to model changes in the data generation mechanism. PF-JMS denotes particle filtering for jump Markov systems, which use sequential Monte Carlo methods to estimate states in a model with evolving probability distributions. KF-BDEMM refers to a Kalman filter-based version of BDEMM; SMC-BDEMM represents a sequential Monte Carlo-based version of BDEMM: GPTS-BDEMM denotes the application of Gaussian process time series models to the BDEMM framework; LDE refers to learning-based dynamic ensembles, which utilize multiple base models with different architectures to improve prediction performance.}\label{fig:timeline}
\end{figure}

In this section, we establish connections between BDEMM and related models and methods in the literature, including Markov switching systems (MSS), GPTS models, forecasting techniques, and dynamic ensemble methods based on learning. Fig.\ref{fig:timeline} depicts a timeline of these works.
\subsection{Connections to MSS}
From a modeling perspective, BDEMM is conceptually related to the MSS or jump Markov system (JMS) model, which dates back to \cite{ackerson1970state,harrison1971bayesian}. There are many variations or extensions of MSS, such as conditional dynamic linear models \cite{west2006bayesian}, dynamic linear models with Markov switching \cite{kim1994dynamic}, the interacting multiple model (IMM) \cite{blom1988interacting,bar1988tracking}, and state-space models with regime-switching \cite{kim1999state}. A set of continuous latent processes comprises state-space models in conjunction with a discrete-time, discrete-state Markov model that captures the unobserved switching process of the prevailing model, used for describing the generative process of the observations in such MSS-type models.

The hierarchical structure of MSS allows them to describe many practical aspects of time series data, such as outliers, nonlinearities, sudden maneuvers in target tracking, and heteroscedasticity. However, most existing MSS approaches use linear Gaussian model components, with some exceptions in e.g. \cite{andrieu2003efficient,chen2000mixture}, which assume linear Gaussian observational models given the current state.

In comparison, BDEMM is more general and expressive in two ways: Firstly, it accommodates both state-space and GPTS model structures, offering greater flexibility to incorporate either linear Gaussian or nonlinear non-Gaussian components for each model mixture element. Secondly, BDEMM offers greater flexibility in expressing model switching processes. It allows the prevailing model to remain unchanged, corresponding to WTT Operator I or to switch over time via WTT Operators II-V. Moreover, different types of model switching like independent switching, Markov-type switching, forgetting mechanism-based switching, and P$\acute{o}$lya urn process-based switching can be performed using these WTT Operators II-V.

From an algorithmic perspective, the KF-based BDEMM algorithm is similar to KF methods applied to the MSS type models, such as the IMM KF \cite{mazor1998interacting}. Two key problems demand specific attention while using them to tackle real-world issues. The first is the approximation of a Gaussian mixture-form posterior distribution by a single Gaussian component at each time step, which can be improved using a parsimonious mixture with a fixed number of Gaussian components to approximate the posterior. The second issue is that the conditional linear system assumption underlying such techniques is problematic to maintain in practice. One may replace routine KFs with the extended KF method to deal with conditional nonlinear systems, as proposed in \cite{chow2013nonlinear}. To address nonlinear non-Gaussian issues generally, SMC methods can be utilized. This extension of the KF-based BDEMM algorithm results in the SMC-based BDEMM algorithm.

The SMC-based BDEMM algorithm has a simple relationship with SMC techniques for MSS, which include the mode (regime) variable into the state vector and apply SMC to this augmented state \cite{mcginnity2000multiple,boers2002hybrid}. Boers et al. proposes an IMM-based PF algorithm that resolves numerical challenges when one mode's probability is significantly low and only a few live particles are allocated to it, while always approximating the posterior via a finite Gaussian mixture, resulting sometimes in loss of accuracy due to accumulated approximation errors over time \cite{boers2003interacting}. In contrast, the SMC-based BDEMM algorithm avoids numerical issues and is not constrained by the finite Gaussian mixture approximation limitation. Recently, both particle Gibbs sampling \cite{el2021particle} and variational inference \cite{khalid2018robust} methods have been implemented in regime-switching state-space models. These works may be considered alternative implementations or algorithmic extensions of the BDEMM framework.
\subsection{Connections to GPTS models}
GPTS models offer probabilistic predictions of future observations as well as estimates of uncertainty. BDEMM utilizes the Bayesian framework inherent within GPTS models and embeds multiple such models for more accurate and robust online prediction.
The resulting INTEL algorithm \cite{liu2020sequential} demonstrates BDEMM's ability to address non-state-space models. Before INTEL's development, most GPTS-based methods required a complicated model structure to capture the potential regime shifts in future observations \cite{chandola2011gaussian,saatcci2010gaussian,roberts2013gaussian,osborne2012real,garnett2010sequential} (e.g., student's t-based observation model in \cite{vanhatalo2009gaussian} or non-stationary kernel function in \cite{garnett2010sequential}, leading to an absence of analytical inference solutions. By using a collection of simple, analytically solvable models instead, the BDEMM framework eliminates the need for complex model specifications, resulting in INTEL having rich modeling capacity to cover complex temporal structures while achieving much higher computational efficiency.
\subsection{Connections to forecasting techniques}
BDEMM has a connection to forecasting theories and techniques that have widespread use in areas such as operations, economics and finance, the energy industry, and environmental research like climate change forecasting. Forecasting methods such as autoregressive integrated moving average (ARIMA) \cite{nelson1998time,box1970distribution}, exponential smoothing (ETS) \cite{chatfield2001new,gardner1985exponential,gardner2006exponential}, etc., assume that the underlying time series is stationary, whereas BDEMM deals with non-stationarity and time-varying uncertainty.

The data preprocessing for time series forecasting includes several practices like Box-Cox transformation \cite{box1964analysis}, time series decomposition \cite{findley1998new,findley2005some,ladiray2012seasonal}, anomaly detection \cite{gupta2013outlier,montero2020fforma}, and feature engineering \cite{stock2002forecasting}. For an extensive review of forecasting models, methods, and applications, see \cite{petropoulos2022forecasting}.

Time series forecasting can be either univariate or multivariate, short-term or long-term, point or interval prediction. The KF-based, SMC-based, GPTS-based BDEMM methods discussed in this paper mainly address univariate, short-term, point forecasting tasks. However, in principle, the BDEMM framework can handle more generic forecasting tasks such as multivariate, long-term, and interval forecasting, subject to the availability of a qualified set of candidate models and an appropriate likelihood function for each model linking its predictions with real observations.
%
%
\subsection{Connections to learning based dynamic ensemble methods}\label{sec:connect_learning}
In recent years, several dynamic multi-model ensemble methods have emerged, including Bayesian optimization (BO) based \cite{du2022bayesian} and reinforcement learning (RL) based methods \cite{fu2022reinforcement}.
Such learning-based dynamic ensemble methods use multiple models with different architectures as the base learners or experts. BDEMM also employs multiple candidate models but does so in a different manner by updating their weights via Bayesian inference.

The BO-based approach evaluates each candidate model's performance based on a validation dataset composed of a batch of recent observations. It also requires another set of recent observations to train the candidate models. Several practical issues need to be addressed, such as determining the sliding window length and dividing it into sub-sets for training and validation. In contrast, BDEMM-based algorithms utilize a set of base models that are not retrained during online prediction. Instead, only the model weights are updated, avoiding the repeated allocation and storage of data for model training. Moreover, BDEMM evaluates the performance of each candidate model differently, measuring its performance as the product of its marginal likelihood and prior probability, without requiring access to a validation dataset.

Furthermore, the BO-based approach employs ten various candidate models, including neural network-based ones, and exploits model diversity that could be beneficial for prediction accuracy but may require large memory space and sacrifice interpretability. In comparison, for BDEMM-based methods, the parameters of each model hold specific physical meanings, and the relationships among the models are clear. While the BO-based method is a pure data-driven approach, BDEMM offers the potential to combine domain knowledge with data to improve predictions.

Dynamic multi-model ensemble or combination can be formulated as a RL problem, as demonstrated in \cite{fu2022reinforcement}. Algorithms are developed to learn a policy for determining the weights of models online, by adapting techniques developed in the RL community. The core of an RL method is usually a Markov decision process (MDP), which is defined as a 4-tuple ($S, A, P_a, R_a$), where $S$ denotes the state space, $A$ denotes the action space, $P_a(s,s')=\mbox{Pr}(s_{t+1}=s'|s_t=s,a_t=a)$ denotes the probability that action $a$ in state $s$ at time $t$ will lead to state $s'$ at time $t+1$, and $R_a(s,s')$ denotes the immediate reward received after transitioning from state $s$ to state $s'$ due to action $a$.

For time series cases considered here, determining the models' weights through action $a_t$ shall not influence the next state value, namely the time series value at time $t+1$. As a result, $P_a(s,s')$ reduces to $\mbox{Pr}(s_{t+1}=s'|s_t=s)$, which represents the state-space model. Hence, RL-based dynamic ensemble techniques naturally connect to BDEMM approaches that employ state-space type models, such as KF-based BDEMM and SMC-based BDEMM presented in Subsections \ref{sec:kf_bdemm} and \ref{sec:smc_bdemm}, respectively.

In principle, RL-based dynamic ensemble techniques inherit the pros and cons of RL itself. The efficacy of RL is sensitive to hyper-parameter values, and selecting appropriate hyper-parameters can be a labor-intensive task \cite{paine2020hyperparameter,paul2019fast,zhang2021importance}. Additionally, typical RL methods may suffer from distribution shift issues \cite{zhao2023context} that frequently occur in non-stationary time series data.

To summarize, BO and RL-based dynamic ensemble techniques are learning-based approaches that train and evaluate multiple models based on a sliding window of recent observations, which assumes a Markovian regime shift process. However, real regime shift processes can be non-Markovian and have a long-term memory structure. It is unclear how to extend these learning-based approaches to fit non-Markovian regime shifts, while BDEMM easily breaks the Markovian assumption by leveraging a non-Markovian WTT operator, such as Operator V described in Subsection \ref{sec:wtt_V}. Compared to BDEMM, these learning-based methods are more data-driven, have more parameters or hyper-parameters to learn, lack interpretability, and require more memory space for training and storing the candidate models.

Both the BO method and BDEMM lie between the two extreme ends of the modeling culture continuum presented in Section 1, where one end is data modeling, and the other is algorithmic modeling
In principle, both learning-based dynamic ensemble techniques and BDEMM lie between the two extreme ends of the modeling culture continuum presented in Section \ref{sec:introduction}, where one end is data modeling, and the other is algorithmic modeling. The main difference between the learning-based method and BDEMM is that the former is closer to the algorithmic modeling end, while the latter is closer to the data modeling end.
\section{Discussions}\label{sec:fut}
In this section, we first discuss when to choose BDEMM over other dynamic ensemble methods and the limitations of the BDEMM framework. We then review how to develop candidate models for BDEMM. Next, we examine two critical computational modules used in BDEMM: the WTT operators and the inference algorithm. Finally, we consider potential future directions for BDEMM.
\subsection{When is it better to select BDEMM instead of other dynamic ensemble techniques?}
Compared to related methods, BDEMM has its own characteristics that make it preferable in certain situations.

Firstly, BDEMM is a direct extension of Bayesian model averaging (BMA) and has more theoretical grounding compared to heuristics-based dynamic ensemble strategies. It inherits all the theoretical properties of BMA, provided an appropriate WTT operator is employed.

Secondly, BDEMM is a probabilistic inference-based framework that does not require access to a validation dataset. The working of BDEMM measures model performance by taking into account both prior probability and marginal likelihood. This reduces reliance on validation datasets and avoids repeated training of candidate models.

Finally, BDEMM provides a flexible and powerful interface that enables the embedding of prior domain knowledge into the algorithm design. This enables the use of pre-trained candidate models and enhances interpretability.

These characteristics make BDEMM preferable to other related techniques, such as the learning-based methods mentioned in Subsection \ref{sec:connect_learning}, when prior domain knowledge about the time series to be investigated exists, available memory space for large candidate models is limited, and interpretability is required.
\subsection{Limitations of the BDEMM framework}\label{sec:limitation}
There are three recognized Bayesian perspectives for model uncertainty quantification: M-closed, M-complete, and M-open. M-closed assumes that the correct model is one of the candidate models being considered, while M-complete and M-open assume that the correct model is not among the candidate models \cite{bernardo2009bayesian}. Traditional BMA theory does not apply to these M-complete and M-open frameworks \cite{clyde2013bayesian,yao2018using}. BDEMM follows the M-closed perspective as a dynamic extension of BMA. Adapting BDEMM to these frameworks remains an open question. One possible approach is to fine-tune or re-learn candidate models from scratch once new observations arrive, which may result in the loss of previously learned knowledge.
This phenomenon, called catastrophic forgetting, has been widely studied in continual learning research \cite{li2017learning,aljundi2018memory,van2018three}. A basic strategy to mitigate catastrophic forgetting is to balance model stability and plasticity \cite{abraham2005memory}.
\subsection{How to Build up Candidate Models for BDEMM?}\label{sec:discuss_model_pool}
BDEMM uses a set of candidate models to capture the uncertainty in the time series. In some cases, the candidate models are already known beforehand, such as in \cite{raftery2010online,liu2011instantaneous}, where each model corresponds to a physical hypothesis.

In other cases, a nominal model may be available by default, which the modeler can adjust or modify to create additional candidate models. For example, in most applications of the KF algorithm, the measurement noise model is assumed to be a zero-mean Gaussian with a fixed variance. If the modeler suspects that outliers may appear, they could build another candidate model by modifying the nominal one, such as increasing its variance value, or modulating its distribution to be heavy-tailed student's t \cite{liu2017robust}.
%
%

In some cases, the modeler may only have access to historical data and no prior knowledge or nominal model. In such situations, a data-driven approach can be used to build candidate models. For example, in \cite{liu2020data}, each candidate model corresponds to a specific segmentation of the historical data, and the parameter values for each candidate model are set via Bayesian optimization to maximize the log-likelihood of the data within that segmentation.

Alternatively, if domain knowledge is available, the modeler can combine it with historical data to build candidate models. One strategy is to first fit a nominal model to the historical data and then build additional candidate models by perturbing the parameters of the nominal model based on domain knowledge. For instance, both the DyEnsemble method \cite{qi2019dynamic} and the INTEL algorithm \cite{liu2020sequential} adopt this perturbation strategy, but differ in their application: DyEnsemble applies stochastic perturbations inspired by the Dropout operation from deep artificial neural networks, while INTEL uses deterministic perturbations.

The BDEMM framework allows for various model structures provided that an appropriate likelihood function can link the model predictions to observed time series data. Examples include state-space models, GPTS models, physical models, data-driven models such as neural networks and support vector regression models, and conventional statistical models like autoregressive moving-average models. The choice of which models to use depends on the modeler's expertise or can be learned from historical data.
\subsection{On WTT Operators}\label{sec:discuss_wtt}
The WTT operator is a crucial component of the BDEMM framework because it determines the prior probabilities for each candidate model at each time instant and connects successive BMA operations. The current version of BDEMM offers five WTT operators that correspond to different assumptions about the model switching law, as described in subsection \ref{sec:wtt}. However, it is up to the modeler to determine beforehand which WTT operator to use for a particular modeling task.
Currently, there is no established methodology for selecting an appropriate WTT operator. Although the current set of operators can cover most realistic cases, there may be instances where they are not suitable. Hence, it is desirable to develop an adaptive approach for WTT operator selection or even design an appropriate operator on-the-fly, especially for novel or complex datasets.
Future research could explore ways to effectively select a suitable WTT operator for specific scenarios, thereby improving the robustness and flexibility of the BDEMM framework.
\subsection{On Inference Algorithms}\label{sec:inference}
Most current BDEMM algorithms use the standard versions of Kalman filters (KF) and sequential Monte Carlo (SMC) as the major inference engine. However, there is potential to combine other types of inference methods with BDEMM, such as variational inference \cite{blei2017variational,zhang2018advances,khalid2018robust}, particle flow \cite{li2017particle}, and approximate Bayesian computation \cite{marin2012approximate,beaumont2009adaptive,sisson2018handbook}.
By combining these advanced inference methods with BDEMM, more robust and accurate online prediction algorithms can be developed for non-stationary time series data. For example, variational inference can approximate complex probability distributions with a simpler distribution, while particle flow can track complex posterior distribution dynamics effectively. Furthermore, approximate Bayesian computation can handle models with intractable likelihood functions and help overcome model misspecification problems.
Future research could explore the integration of these advanced inference methods and BDEMM, potentially yielding more powerful and versatile tools for predictive modeling with non-stationary time series data.
\subsection{On Other Future Research Directions}
Except those mentioned in the above subsections, here we suggest several other potential directions for future research following the BDEMM framework:
\begin{itemize}
  \item Develop learning assisted BDEMM to overcome the limitations of BDEMM in handling M-complete and M-open cases, as discussed in Subsection \ref{sec:limitation}. By incorporating machine learning techniques, such as deep learning and reinforcement learning, it may be possible to develop more robust and flexible BDEMM models and algorithms.
  \item Investigate the potential of the BDEMM framework for online continual learning problems, especially the stability-plasticity dilemma \cite{jung2023new}. This direction may focus on developing techniques that allow BDEMM to adapt to changing data distributions while retaining previously learned knowledge.
  \item Analyze BDEMM from a loss-theoretic perspective, which is commonly used in the machine learning community. A loss-theoretic analysis may provide insights into the generalization properties of BDEMM models and algorithms and lead to improved theoretical guarantees for their performance.
  \item Develop likelihood-free BDEMM models and algorithms, following the spirit of generalized Bayesian inference \cite{boustati2020generalised,zellner1988optimal,bissiri2016general}. This direction may focus on developing techniques that can handle models with intractable likelihood functions or are otherwise challenging to model using traditional Bayesian methods.
\end{itemize}
\section{Open Resources}\label{sec:open}
Here we introduce open-source codes that can be used for implementing BDEMM-type algorithms and some benchmark time series datasets.
\subsection{Open-Source Codes}
Open-source codes have been developed for RPF and RMMDF, as described in subsections \ref{sec:rpf} and \ref{sec:mmdf}, available at \url{https://github.com/robinlau1981/dmapf} and \url{https://github.com/robinlau1981/fusion}. The code related to the toy experiment presented in subsection \ref{sec:toy_exp} is available at \url{https://github.com/robinlau1981/BDEMM}.

In Section \ref{sec:theory}, importance sampling is discussed and its key role in estimating marginal likelihoods of models highlighted. Advanced importance sampling methods like AAIS (adaptive annealed importance sampling) \cite{liu2014adaptive} are helpful for multimodal posterior exploration. Matlab codes for AAIS can be obtained from \url{https://github.com/robinlau1981/MarginalLikelihoodEstimator} and \url{https://github.com/robinlau1981/AAIS}. Layered AIS \cite{martino2017layered} can be implemented using the code at \url{http://www.lucamartino.altervista.org/CODE_LAIS_v03.zip}.

For SMC, Nando de Freitas' Matlab codes are recommended and can be accessed from \url{http://www.cs.ubc.ca/~nando/software/upf_demos.tar.gz}. Similarly, Nicolas Chopin has an SMC Python library located at \url{https://nchopin.github.io/software/} that implements different versions of SMC described in \cite{chopin2020introduction}.

A resource pool for GPs that includes a large volume of software, papers, books, and events can be found at \url{http://www.gaussianprocess.org/}.
%
%
%
\subsection{Benchmark Time Series Datasets}
There exist some commonly used datasets for time series forecasting or anomaly detection, which researchers and practitioners often use to benchmark their methods. These include:
\begin{enumerate}
  \item The M-competition dataset, which is available at \url{https://forecasters.org/resources/time-series-data/}. It contains various types of time series data, such as macroeconomic indicators, financial indices, and demographic data.
  \item The PEMS dataset, which can be accessed at \url{https://archive.ics.uci.edu/ml/datasets/PEMS-SF}. This dataset includes traffic flow data from San Francisco Bay Area highways.
  \item The Numenta dataset \cite{lavin2015evaluating}, which comprises both synthetic and real-world data and aims to evaluate anomaly detection methods on streaming time series data.
  \item The Amazon’s CPU usage dataset \cite{ahmad2017unsupervised}, which contains time series data for CPU usage in cloud servers.
  \item The well-log dataset \cite{turner2009adaptive,turner2012gaussian}, which includes time series data collected from oil wells.
  \item The fish killer dataset \cite{liu2020sequential}, which contains environmental and operational data from a fish farm.
  \item The ``30 industry portfolios" dataset \cite{xuan2007modeling}, which consists of monthly returns on portfolios of stocks from different industries.
\end{enumerate}
For more datasets used in time series forecasting, readers are referred to the appendix section of \cite{petropoulos2022forecasting}.
\section{Conclusions}\label{sec:conc}
BDEMM is a dynamic extension of BMA theory that has found successful applications in sequential online prediction problems. This paper provides a comprehensive introduction to BDEMM, including its theoretical foundations, algorithms, practical applications, connections to other research, and a discussion of its strengths, limitations, and potential future directions.

It is important to note that the topics related to sequential online prediction and model ensemble are broad, so this paper focuses on the most representative and relevant works with respect to BDEMM based on personal evaluation. It is hoped that this effort will stimulate more theoretical and applied research on BDEMM, as it has significant potential to advance predictive modeling in fields such as finance, ecology, engineering, and geology, among others.
\bibliographystyle{IEEEbib}
\bibliography{mybibliography}
\end{document}